\documentstyle[11pt,aasms4]{article}
\doublespace{\parskip 4pt plus 1.5pt
\baselineskip 26pt plus 1pt minus .5pt
\lineskip 6pt plus 1pt \lineskiplimit 5pt }
\def\beq{\begin{equation}}
\def\eeq{\end{equation}}
\def\beqar{\begin{eqnarray}}
\def\eeqar{\end{eqnarray}}

\newcommand{\curl}{{\nabla \times}}

\newcommand{\p}{{\partial}}

\newcommand{\gapprox}{\lower.4ex\hbox{$\;\buildrel
>\over{\scriptstyle\sim}\;$}}
\newcommand{\lapprox}{\lower.4ex\hbox{$\;\buildrel
<\over{\scriptstyle\sim}\;$}}

\lefthead{Kim, Diamond, \& Zweibel}
\righthead{ambipolar diffusion}

\begin{document}
\def\ref#1{\hangindent=6em\hangafter=1 {#1}}

\title{TURBULENT DIFFUSION OF MAGNETIC FIELDS IN WEAKLY IONIZED GAS}
\author{EUN-JIN KIM and P.H. DIAMOND}

\affil{Department of Physics,
University of California, San Diego,
La Jolla, CA 92093-0319}

\begin{abstract}
The diffusion of uni-directional magnetic fields by 
two dimensional turbulent flows in a weakly ionized gas is studied.
The fields here are orthogonal to the plane of fluid motion.
This simple model arises in the context of the decay of the mean
magnetic flux to mass ratio in the interstellar medium.
When ions are strongly coupled to neutrals, the transport of a large--scale
magnetic field is driven by both turbulent mixing and nonlinear, ambipolar drift.
Using a standard homogeneous and Gaussian statistical model for turbulence,
we show rigorously that a large-scale magnetic field can decay
on at most turbulent mixing time scales when the field and neutral flow are strongly
coupled. There is no enhancement of the decay rate by ambipolar diffusion.
These results extend the Zeldovich theorem to encompass the regime
of two dimensional flows and orthogonal magnetic fields,
recently considered by Zweibel (2002). The limitation of the strong coupling
approximation and its implications are discussed.

{\it Subject headings: ISM: magnetic fields --- magnetic fields --- turbulence --- diffusion} 

\end{abstract}
\pagebreak

\section{INTRODUCTION}

In the interstellar medium, where the bulk of fluid consists of neutral gas,
magnetic fields appear to be constantly lost from the 
fluid---a phenomenon often referred to as ambipolar diffusion (Spitzer 1978). 
It is simply because magnetic fields move
with (or are tied to) ionized gases while there is slippage between the motion
of ionized and neutral gases. 
For parameter values typical of interstellar clouds, the ambipolar drift,
however, appears to be too slow (roughly by two orders of magnitude)
to explain the large dispersion in the correlation
between density and magnetic field strength (Zweibel 2002). 
This naturally motivates the exploration of other transport processes.
In particular, the interaction of nonlinear, ambipolar diffusion and turbulent
advective mixing is a question of obvious relevance. 
In this Letter, we seek
to examine the interplay of these two processes and to determine a bound
on the turbulent transport of magnetic fields. 

While it is well known that turbulent mixing leads to a rapid diffusion of  
passive scalar fields, this is no longer 
the case for the diffusion of even a weak magnetic field 
(far below equipartition) in a fully ionized gas (Cattaneo \& Vainshtein 1991;
Gruzinov \& Diamond 1994), due to the backreaction of the Lorenz force. 
In a weakly ionized gas, such as in the interstellar medium, the
problem becomes more complicated since turbulent mixing also depends on
the collision frequency between ions and neutrals as well as on 
the strength of magnetic fields. It is because in  a weakly ionized gas,
ions undergo the frictional damping due to collisions with neutrals, which
effectively reduces the effect of the Lorentz force (or Alfv\'en waves). 
In fact, the previous work by Kim (1997) demonstrated that in two spatial
dimensions (2D), the diffusion is still reduced below its kinematic value, but the 
critical strength of a large--scale magnetic field above which the diffusion is 
reduced can be larger (by a factor of $\sqrt{\nu_{in}\tau}$, where $\nu_{in}$ and $\tau$ 
are the ion-neutral collision frequency and correlation time of neutrals) 
as compared to what happens in the case of a fully ionized gas. 

For simplicity, in this Letter, we consider the mixing of uni-directional 
magnetic fields by two dimensional flows which are perpendicular to these fields. 
Note that this configuration is different from traditional 2D MHD, where the
fields and flows are coplanar.
To maintain this geometry, it is necessary to make the strong coupling 
approximation (Spitzer 1978; Shu 1983; Zweibel 1988), by assuming 
that the drift between ions and neutrals is balanced by the Lorentz force 
on ions due to frequent ion-neutral collisions (frictional damping). 
Our work is the generalization of Zweibel (2002) who considered the diffusion of 
magnetic fields by highly idealized flows made up of an ensemble
of hyperbolic stagnation points. Since it is not altogether clear how to relate these
flows to realistic turbulence models, we take a statistical approach here and
rigorously derive the diffusion rate by assuming a standard scenario of Gaussian 
and homogeneous turbulence.
Note that it is possible that this simplified statistical model may fail 
when the nonlinear, ambipolar diffusion is dominant, in which sharp front-like
structures are generated (Brandenburg \& Zweibel 1994). 

Under the strong coupling approximation, magnetic fields are advected passively
by neutral flows and diffused by nonlinear ambipolar drift (in addition to the
usual Ohmic diffusion). 
Thus, in view of this nonlinear diffusion, one may naively
expect that a large--scale magnetic field would decay at a rate which is significantly
enhanced over the turbulent (kinematic) value. We show, however, that it is
not the case, because of strong fluctuations (see \S 3).  
Specifically, in deriving a generalized Zeldovich theorem 
(which relates the macroscopic quantity (transport) to microscopic dissipation)
for a weakly ionized gas in 2D motion but containing magnetic fields orthogonal
to the plane of motion, 
we show that the flux transport due to the advection by neutral flows has an upper
bound given by kinematic value while the nonlinear diffusion arising from
ambipolar drift is insignificant. The remainder of the Letter is
organized as follows. We present our model in \S 2 and derive a diffusion rate
in \S 3. Section 4 contains the summary and discussion of the Letter.

\section{MODEL}

In a weakly ionized gas with $\rho \sim \rho_n \gg \rho_i$, $\nu_{in}/\nu_{ni}$
is large, i.e., $\nu_{in}/\nu_{ni} = \rho_n/\rho_i \gg 1$. Here, $\rho$, $\rho_n$, 
and $\rho_i$ are the density of
the bulk of fluid, neutrals, and ions; and $\nu_{in}$ and $\nu_{ni}$ are
ion-neutral and neutral-ion collision frequencies, respectively.
Infrequent neutral-ion collisions permit us to prescribe the motion of
neutral gas, provided that
\begin{eqnarray}
{B^2 \over 4 \pi \rho_i} {\nu_{in}\over \nu_{ni}} \hskip0.2cm (\simeq
v_A^2) \ll    v^2\,,
&& \tau \nu_{ni}  \ll 1\,,
\label{eq1}
\end{eqnarray}
where $v\sim v_n$ and $\tau$ are the neutral velocity and correlation time,
and  $v_A = B/\sqrt{4 \pi \rho}$ is the Alfv\'en speed with respect to the bulk of the 
fluid (see Kim 1997). 
Given a neutral velocity, the ion velocity follows simply from the strong coupling approximation 
as ${\bf v}_i = {\bf v}_n + [(\curl {\bf B}) \times {\bf B}]/4 \pi \rho_i \nu_{in}$.
Note here that the strong coupling approximation is valid when 
\begin{eqnarray}
\nu_{in}&>& {\tilde \nu}_{A}\,, 
\label{eq2}
\end{eqnarray}
where ${\tilde \nu}_A = k B/\sqrt{4 \pi \rho_i} = k v_A \sqrt{\nu_{in}/\nu_{ni}}$
is the Alfv\'en frequency defined by using the density of ion (cf. equation [1]),
with $k$ being wave number. 
When this condition is violated, as is likely to
be case on small scales due to the high frequency of Alfv\'en waves, 
the drift between ion and neutral motions is no longer balanced by
Lorentz force, thereby requiring a self consistent treatment of ion dynamics
(Kim 1997).

We consider the mixing of uni-directional magnetic fields (say, ${\bf B} = B(x,y) {\hat z}$)
by incompressible neutral flows ${\bf v}(x,y)$ in the $x$-$y$ plane perpendicular to these fields.
Thus, we treat the 2D motions of 3D fields.
Note that our problem is somewhat similar to that in the Goldreich-Sridhar model (Goldreich and Sridhar
1997), which also considers perpendicular mixing of anisotropic structures.
The evolution equation for the strength of magnetic field in this geometry can be written in 
the following form (see also Zweibel 2002):
\begin{eqnarray}
(\p_t  + {\bf v} \cdot \nabla) Q 
&=& \alpha \nabla \cdot [Q^2 \nabla Q]\,.
\label{eq3}
\end{eqnarray}
Here, $Q \equiv B/\rho$ and $\alpha \equiv \rho /4 \pi \nu_{ni}$, and the (small) Ohmic
diffusivity has been ignored. 
The second term on the left hand side of equation (2) represents
the advection by neutral flow while the term on the right hand side is
the nonlinear diffusion by ambipolar drift. The ratio of the effects of
these two can be measured by ambipolar Reynolds number $R_{AD} = vl/\lambda_T
= (v^2/v_A^2) (\tau \nu_{ni})$.
Here, $vl = \eta_k$ is the kinematic diffusion rate, and 
$\lambda_T = v_A^2/ \nu_{ni}$ is 
ambipolar drift due to total magnetic fields. 
Note that $R_{AD}$ can be larger or smaller than unity, while still 
being consistent with equation (1).  
The question is then what the total diffusion rate is in the presence of these two 
(advection and ambipolar drift) effects. Would they act together to significantly
enhance the diffusion rate over the kinematic value $\eta_k$? 
To answer this question, we assume Gaussian, homogeneous turbulence 
and evaluate the diffusion rate by using a quasi-linear or `second order
smoothing' closure (for instance, see Moffatt 1978) in the next section. 
The observant reader is no doubt puzzled by the fact that equation (3)
is two dimensional. The motivation for this simplification 
is that turbulence with ${\bf k} \cdot {\bf B}_0 \ne 0$ will bend
magnetic field lines, resulting in its conversion to Alfv\'en waves
and radiation along ${\bf B}_0$. Such fluctuations are intrinsically
less effective at transporting $Q$, as much of their energy is
expended on bending. However, flute-like eddys, with 
${\bf k} \cdot {\bf B}_0=0$, are energetically favored for transport,
and also remain correlated with the flux tube being transported for
a longer time. 
Thus, the incorporation of eddys with finite wavenumber along ${\bf B}_0$ 
(i.e., ${\bf k} \cdot {\bf B}_0\ne 0$) will reduce transport of magnetic
fields.

\section{Diffusion rate}

We employ the decomposition of $Q$ into  large--scale $\langle Q \rangle$ 
and small--scale components $Q'$ and assume that flows are on small--scales
($\langle {\bf v} \rangle = 0$). The equations for $\langle Q \rangle$ and $Q'$
are then easily obtained as follows:
\begin{eqnarray}
\p_t \langle Q \rangle  + \nabla \cdot \langle {\bf v} Q'\rangle 
&=& \alpha \nabla \cdot \langle Q^2 \nabla Q\rangle\,,
\label{eq4}\\
\p_t Q' + {\bf v} \cdot \nabla \langle Q\rangle
&=& \alpha \nabla \cdot {\bf F }\,,
\label{eq5}
\end{eqnarray}
where 
\begin{eqnarray}
{\bf F} &\equiv& [Q'^2-\langle Q'^2 \rangle ] \nabla \langle Q \rangle
+ 2 Q' \langle Q \rangle \nabla \langle Q \rangle 
+ \langle Q \rangle^2 \nabla Q'
\nonumber \\
&& + Q'^2 \nabla Q' - \langle Q'^2 \nabla Q' \rangle
+ 2\langle Q \rangle [Q' \nabla Q' - \langle Q' \nabla Q'\rangle ]\,.
\nonumber 
\end{eqnarray}
As can be seen from equation (4), the determination of the diffusion rate  requires 
the computation of the flux $\Gamma_i = \langle { v}_i Q'\rangle$ and 
the cubic nonlinear term in $Q$. To compute the cubic nonlinear 
term, as well as other nonlinear terms that appear in the following
analysis, we assume that the statistics of fluctuations are 
Gaussian and that the turbulence is homogeneous.
Then,
\begin{equation}
\langle Q^2 \nabla Q\rangle
= [\langle Q \rangle^2 + \langle Q'^2 \rangle] \nabla\langle Q\rangle\,.
\label{eq6}
\end{equation}
On the other hand, the flux $\Gamma_i$ is evaluated 
by assuming stationary turbulence. We first 
multiply equation (5) by $v_i$, and then take the average to obtain, 
\begin{eqnarray}
\Gamma_i &=& -{\eta_k\over 1 + \alpha \tau k_{eff}^2 \langle Q^2 \rangle} 
\p_i \langle Q \rangle
= -{\eta_k\over 1 + R_{AD}^{-1}} \p_i \langle Q \rangle\,.
\label{eq7}
\end{eqnarray}
Here, $\eta_k = \tau \langle v^2 \rangle/2 \simeq vl$ is the kinematic diffusion
rate; $k_{eff} \simeq 1/l$ is the inverse of the characteristic scale of fluctuating
magnetic fields; $R_{AD} = \eta_k/\lambda_T$ where 
$\lambda_T = [\langle v_A\rangle ^2 +\langle v_A'^2 \rangle ]/\nu_{ni}$;
$\tau$ is the correlation time of fluctuating magnetic fields, which is assumed
to be comparable to that of neutral velocity. 
Note that $\lambda_T$ and $R_{AD}$, now defined in terms of averaged quantities, include
both mean and fluctuating components.
In deriving equation (7), we used $\langle v Q'^2 \rangle = 0$ and 
$\langle v_i v_j \rangle = \delta_{ij} \langle v^2 \rangle /2$ by assuming
an isotropic turbulence.
Equation (7) states how much flux is transported from large to small scales,
thereby leading to the decay of  $\langle Q \rangle$. Interestingly, 
the diffusion rate by advection, $-\Gamma_i/\p_i \langle Q \rangle$,
has an upper bound given by kinematic diffusion $\eta_k$,
and becomes smaller as $R_{AD}$ decreases. This is because that ambipolar drift 
`renormalizes' the correlation time so as to reduce the transport. 
Thus, it is clear that the kinematic turbulent flux is an \underline{upper}
\underline{bound} on $\Gamma_i$.  

In the case of stationary turbulence, the flux transport  
is balanced by dissipation on small scales as:  
\begin{eqnarray}
\Gamma_i \partial_i \langle Q \rangle
&=& -\alpha [ \langle Q \rangle^2 + \langle Q'^2 \rangle]
\langle (\partial_i Q')^2 \rangle
= -\lambda_T \langle (\partial_i Q')^2 \rangle
\,.
\label{eq8}
\end{eqnarray} 
This was obtained by multiplying 
equation (5) by $Q'$ and then taking the average. 
Equation (8),  together with (7), establishes the relation between small and large scale fields as
\begin{eqnarray}
{\eta_k\over 1 + R_{AD}^{-1} } 
(\p_i \langle Q \rangle)^2
&=& \lambda_T 
\langle (\partial_i Q')^2 \rangle\,.
\label{eq9}
\end{eqnarray} 
Equation (9) is a generalized Zeldovich theorem for a weakly ionized, strongly coupled
gas. It gives the relation between mean field and its gradient and
relates the macroscopic quantity (i.e., flux transport) to microscopic
dissipation. 
Note that
the original Zeldovich theorem in a fully ionized gas can be recovered by replacing 
the ambipolar drift by Ohmic diffusion $\eta$ ($\lambda_T \to  \eta$) and by
taking $R_{AD}^{-1}=0$, which gives 
$ \langle (\partial_i Q')^2 \rangle/(\p_i \langle Q \rangle)^2 = \eta_k/\eta = R_m$
($R_m$ is the magnetic Reynolds number).
Of course, the situation discussed here is 3D, with fields orthogonal to the
plane of 2D motion.

Finally, the (total) diffusion rate of $\langle Q \rangle$ follows
from equations (5)---(7) and (9) as
\begin{eqnarray}
\eta_T &=& 
{\eta_k\over 1 + R_{AD}^{-1}} 
\left[ 1 + 
{ |\nabla \langle Q \rangle |^2 \over \langle (\nabla Q')^2 \rangle } \right]\,.
\label{eq10}
\end{eqnarray} 
$\eta_T$ given in this form illustrates the two complementary effects of ambipolar
drift on diffusion---the first is the reduction of $\eta_T$ by renormalization of $\tau$
(the term $R_{AD}^{-1}$) and the second the enhancement of $\eta_T$ by nonlinear diffusion
(the second term in the square brackets). 
Due to the second effect, it is, in principle, possible that $\eta_T \gg \eta_k$.
This, however, turns out to be very unlikely. To see this,
we first estimate $\eta_T$ in an interesting and more relevant case where $R_{AD}>1$. 

When $R_{AD}>1$, equation (9) leads to 
\begin{eqnarray}
{ \langle (\nabla Q')^2 \rangle \over |\nabla \langle Q \rangle |^2  } 
&\sim &   R_{AD} \,\,\,(> 1)\,,
\label{eq11}
\end{eqnarray}
suggesting strong fluctuations. 
Equation (10) then becomes 
\begin{equation}
\eta_T \sim \eta_k\,.
\label{eq12}
\end{equation}
Thus, $\eta_T$ approaches
the kinematic value $\eta_k$. That is, the diffusion cannot exceed the kinematic
rate, because of strong fluctuations.
Note that in this limit, the field is transported as an effectively passive scalars!

We now look at a less interesting limit  $R_{AD} < 1$ where 
the effect of turbulence does not play an important role.
Note that in this limit, the strong coupling approximation
can easily break down. This can be seen by rewriting
the validity condition for the  strong coupling approximation (equation [2])
as $R_{AD}> 1/(\tau \nu_{in})$. That is, 
the smaller $R_{AD}$ (the stronger magnetic field), the easier it is to violate the 
strong coupling approximation. Thus, the results obtained in the limit $R_{AD}\ll 1$ 
may not be consistent with this approximation. With this in mind, 
we reduce equation (9) to 
\begin{eqnarray}
{ \langle (\nabla Q')^2 \rangle \over |\nabla \langle Q \rangle |^2  } 
&\sim &   R_{AD}^2\,\,\, (< 1)\,,
\label{eq13}
\end{eqnarray}
for $R_{AD}<1$. 
Equation (13) indicates that to satisfy the stationarity condition (equation [9]),
$\langle Q\rangle ^2 > \langle Q'^2 \rangle$. This follows because the dissipation
due to ambipolar drift on small scales is too large to be balanced by flux transport. 
However, as noted previously, the strong coupling approximation (which leads to the nonlinear 
diffusion) is likely to be invalid on small scales, especially when magnetic fields
are strong. What should happen on small scales is propagation of
Alfv\'en waves, rather than nonlinear diffusion. 
Alternatively put, 
when the full dynamics of ions is taken into account, stationary 
turbulence may still be possible even when $\langle Q\rangle ^2 > \langle Q'^2 \rangle$.
Now, the diffusion rate in this case  ($R_{AD}<1$) follows from equations (10) and (13) as
\begin{equation}
\eta_T  \sim \eta_k/R_{AD} = \lambda_T \simeq 
\langle v_A \rangle^2 /\nu_{ni}\,.
\label{eq14} 
\end{equation}
This is a somewhat expected result in the sense that for small $R_{AD}$,
the diffusion rate is set by the ambipolar drift. Our non-trivial result is 
the observation that in this case, though the ambipolar drift due to fluctuating 
magnetic fields must be negligible in order to maintain stationarity.  

\section{SUMMARY AND DISCUSSION}

The problem of transport of magnetic fields in the
interstellar medium is studied by incorporating the effect of turbulence.
Specifically, we
consider the diffusion of uni-directional magnetic fields in
the presence of 2D incompressible, turbulent (neutral) flows perpendicular to these magnetic fields,
embedded in a weakly ionized gas. By assuming that the strong coupling approximation
is valid on all scales, we compute the total diffusion rate of a large--scale magnetic 
field through a quasi-linear analysis. 
When the
turbulence is homogeneous, stationary, and Gaussian, the diffusion rate $\eta_T$
is found to depend on $R_{AD}$ and the level of fluctuations (see equation [10]),
with ambipolar drift playing two complementary roles (see \S 3).
In particular, when  $R_{AD}>1$, $\eta_T$ is shown to be at most of order the 
turbulent rate $\eta_k = vl$. 
In this case, the field is effectively a passive scalar.
Interestingly, this suggests that even in the strong coupling regime,
it is unlikely that magnetic fields will  diffuse at a rate faster than the simple
kinematic value, in spite of the nonlinear diffusion operator.
In the opposite case ($R_{AD} < 1$), we demonstrated $\eta_T \simeq
\lambda_T$ as long as fluctuations are negligible compared to the mean field.
Note, however, that this limit may not be consistent with the strong coupling approximation.  
Therefore, our result not only confirms the main point of Zweibel (2002)
but also puts it on a simple, rigorous foundation.

The results of this Letter are applicable to any 
system with a neutral populations and weak magnetic field, such that the 
strong coupling approximation (${\bf v}_i = {\bf v}_n + [(\curl {\bf B}) \times {\bf B}]/4 \pi
\rho_i \nu_{in}$) is valid. 
However, because of the assumed incompressibility of neutral flows, a more uniform
loading of 
magnetic fields discussed in the paper is basically due to
the diffusion of magnetic fields in a constant density background.
As the magnetic field diffuses while the density remains constant,
system progresses toward a state of more uniformly loaded magnetic fields.
In the turbulent case with $R_{AD}>1$, this mass loading uniformization occurs
by turbulent cascade of magnetic energy (by diffusive mixing) to small scale,
where it is eliminated by Ohmic dissipation.
This uniformization occurs in one large eddy turnover time, as one expects,
since $\tau_{un} = \sum_n \tau_n = \sum_n{l_n/ v_n} = \sum_n (l_n/
\epsilon^{1/3} l_n^{1/3}) 
= (l_0^{2/3}/\epsilon^{1/3})(1/( 1-\alpha^{2/3}))
\sim \tau_0$. Here, $\epsilon$ is the energy dissipation
rate, and 
$l_n = \alpha \l_{n-1}$ with $\alpha \sim 1/2$ was used.
It is important to realize that the breakdown of flux freezing on small
scales due to Ohmic diffusion is critical to the uniformization of loading.
Strictly speaking, our results cannot be directly applied to
star forming regions, where the compressibility of flows and gravity are crucial.
Nevertheless, our results imply that the ambipolar drift in turbulent
medium can make magnetically subcritical clouds supercritical and also
that the turbulent mixing can uniformize the loading of magnetic field lines on a large
eddy turnover time scale. 

In the interstellar medium, 
the diffusion rate due to ambipolar drift alone is too small (by a few orders of magnitude) 
to explain the observations. Given that $R_{AD}$ based on a large--scale magnetic field
seems to be larger than unity, turbulent mixing perhaps provides a mechanism
by which uniformity of the density and the strength of magnetic fields 
is achieved on the eddy turnover time scale. 
However, ions and neutrals are unlikely to be strongly coupled on small
scales due to the high frequencies of Alfv\'en waves, thereby invalidating
the strong coupling approximation. Therefore, the complete answer 
to the problem ultimately requires the self--consistent 
treatment of ion dynamics. 
Furthermore, for a better estimate on $R_{AD}$, some information on
the strength of fluctuating magnetic fields is needed. For instance,
when fluctuations are much stronger than mean fields, $R_{AD}$ based
on fluctuations may be smaller than the unity, and thus ambipolar drift alone
may lead to a fast diffusion. 
Of course, even in this case, the validity of the nonlinear (ambipolar)
diffusion may become questionable due to the breakdown of the strong coupling 
approximation. In either cases,
the relaxation of strong coupling approximation is expected to bring in the 
reduction of the diffusion
as magnetic fields are no longer passively advected/distorted (Kim 1997). 
Of course, other effects, such as gravity, the detailed microscale mechanism
for dissipating magnetic energy, and turbulence intermittency must be considered
as well. A detailed study of this issue will be addressed in a future work.

\vskip1cm

We thank E.G. Zweibel for helpful comments. 
EK and PHD are supported by the U.S. Department of Energy under Grant No. FG03-88ER 53275.

\end{document}